\documentclass[twocolumn,twoside,slac_two]{revtex4}
\usepackage{graphicx}
\usepackage{fancyhdr}
\begin{document}

\newcommand{\diffunit}{$\mathrm{GeV\;cm^{-2}\;s^{-1}\;sr^{-1}}$}
\newcommand{\pointunit}{$\mathrm{TeV\;cm^{-2}\;s^{-1}}$}
\newcommand{\dNdE}{E^{2}_{\nu} \times dN_{\nu}/dE_{\nu}}
\newcommand{\Nch}{$N_{\mathrm{ch}}\;$}
\newcommand{\ea}{{\it et al} }
\newcommand{\ic}{IceCube}
\newcommand{\esqdnde}{$\mathrm{E^{2}_{\nu} \times dN_{\nu}/dE_{\nu}}$}
\newcommand{\puneicrc}{2005 Proc. 29th Int. Cosmic Ray Conf., Pune}
\newcommand{\ar}{Ahrens J {\it et al} }
\newcommand{\am}{Ackermann M {\it et al} }
\newcommand{\ab}{Achterberg A {\it et al} }

\title{The search for extra-terrestrial sources of high energy neutrinos}


\author{Gary C Hill}
\affiliation{University of Wisconsin, Madison}

\begin{abstract}
The field of high-energy neutrino astronomy has seen rapid progress over the last 15 years, with the
development and operation of the first large-volume detectors. Here, we review the motivation
for construction of these large instruments and discuss what construction and physics
progress has been made. 
\end{abstract}

\maketitle

\thispagestyle{fancy}


\section{Why neutrino astronomy?}
We live in a mysterious universe -- one that abounds with many objects that seem to
involve extemely high-energy processes -- accretion of matter into
black holes at the centres of active galaxies, supernovae and 
gamma-ray bursts, where enormous amounts of energy are released over time scales
as short as a few seconds. Understanding these objects and the processes therein involves
observing high-energy radiation and particles.
 Our three particle messengers we have for high-energy 
astronomy are charged cosmic rays (protons and nuclei), gamma-rays and neutrinos. 
   While we have large amounts of data from cosmic ray and gamma-ray observations, the
nature of their sources is still not completely understood. It is the 
neutrino that may provide the connection.

The cosmic rays -- high energy protons and nuclei -- have been well studied at earth with both
space and ground-based detectors. Their major astronomical disadvantage is that they are charged,
and thus except for the very highest energy protons,
 they spiral around in magnetic fields during their passage to earth, which means that
knowledge of their original direction is lost. High-energy gamma-rays have been detected from
many galactic and extra-galactic objects but their usefulness as a messenger over long cosmic
scales is limited by their absorption on the extra-galactic background light.
                                                                                                                        
The Auger cosmic ray detector\cite{auger} in Argentina is the culmination of nearly a century
of effort studying the high energy cosmic ray particles. The array combines for the first
time the two key techniques used over recent decades - a ground based air shower
particle detector and a series of air fluorescence detectors, which observe light
emitted as the air showers develop in the atmosphere.
The combination of the two techniques has provided an essential energy calibration
for the ground array. The Auger experiment has published a correlation analysis which
hints that the highest energy cosmic rays might be associated with nearby active galaxies\cite{auger-science}.
                                                                                                                        
The high energy gamma-rays are detected using large area ground based
telesopes, which either image the Cerenkov light released from
air showers created by the interaction of the gamma-rays in the atmosphere, or
measure the arrival of the shower particles at the ground. The
Milagro detector\cite{milagro} uses a large pool of water to measure the shower
 particles at the ground. It has a large
sky coverage and has successfully detected sources of high energy gamma rays, including
several in the galactic plane and one very bright source in the Cygnus region\cite{Milagro-Cygnus}.
The HESS\cite{HESS}, MAGIC\cite{MAGIC} and VERITAS\cite{VERITAS} telescopes 
image the shower Cerenkov light pool directly
 and have observed many sources, 
 both galactic and extra-galactic\cite{TeVCat}.
                                                                                                                        
If one of these gamma-ray sources was found to   be a neutrino source, then
a hadronic accelerator central engine might be simultaneously driving cosmic ray, gamma and neutrino
production from the one object\cite{PR}.

The road to a kilometre-scale neutrino detector, pioneered by the DUMAND 
collaboration, has seen the operation of the first generation experiments,
AMANDA and Lake Baikal, as well as initial construction and planning for IceCube,
ANTARES, NESTOR, NEMO and KM3NET. 
The discovery  of neutrinos with these detectors will  hopefully  extend and complement the
knowledge of the universe to date gained through cosmic ray and gamma ray observations.

% If one of these turned out to also be a neutrino source, then
% hadronic acceleration might be the key to the puzzle of what
% central engine might be driving cosmic ray, gamma and neutrino
% production\cite{PR}.

A large volume neutrino detector uses an array of photomultipliers to
record Cherenkov light from through-going muons, or from point-like shower (``cascade'')
events. Muons result from charged current interactions of neutrinos in the detector 
volume, or in the surrounding ice and rock. Cascade events result from charged and
neutral current interactions of all neutrino flavours. 

The backgrounds to a search for a flux of high-energy extra-terrestrial neutrinos
at the earth are atmospheric muons and neutrinos from the interaction of cosmic rays in the 
earth's atmosphere. 
 The atmospheric muons are eliminated by looking for events moving
upward through the detector -- only neutrinos can penetrate the earth. A small fraction of
the large downgoing muon  flux  will be falsely reconstructed in the upward direction.
These are removed by  tight requirements on the fitted track - where only the most neutrino
like events are kept. 
  After atmospheric muons are eliminated, there is a flux of atmospheric
neutrinos
 seen in a detector. This can be used as a calibration test beam to check
the understanding of the detector, or be used to look for new neutrino physics. 
 A search for point sources of neutrinos is made by looking for an excess of events from
a direction in the sky. Electromagnetic observations by other 
 detectors may provide information to reduce the
time over which such a search is made - for instance in a search for neutrinos
correlated with  gamma-ray bursts. 
 One can also look for a diffuse excess of neutrinos from the sum of all sources in the 
universe. Since the extra-terrestrial flux predictions tend to go as $dN/dE \sim E^{-2}$,
 one 
looks for higher energy events in the detector to separate them from the more
steep atmospheric neutrino spectrum ($dN/dE \sim E^{-3.7}$).

\section{Northern-hemisphere detectors}
The Baikal collaboration has constructed a neutrino detector in the deep fresh water lake Baikal
in Siberia\cite{Baikal98-03}. In the winter, the surface of the lake freezes over allowing the team to move
equipment out to the site and lower the strings of detectors to the bottom of the lake. The
detector has operated for many years and has produced limits of the fluxes of diffuse neutrino
sources.
 In the energy range ($20-5\times10^{4}$ TeV),
 the Baikal collaboration has
analysed  1038 days (1998-2003) of
data from  the NT-200 experiment, leading to a limit on a diffuse flux of neutrinos
from the sum of all sources in the universe of
 \esqdnde $= 8.1\times10^{-7}$ \diffunit \cite{Baikal98-03}.

The ANTARES detector\cite{ANTARES}, located in the deep Mediterranian ocean, has been 
recently completed\cite{ANTARES-nu2008}. This consists of 12 lines, a total of 900 optical modules. The
optical modules are deployed in triplets (a ``storey'') spaced at 14.5 metres. 
 Deployment of each line is done from a ship, which lowers the line to the
ocean bottom. After that is completed, a miniature submarine is used to complete
the electrical connection of the line into a junction box, from where data is
transmitted back to a shore station via an undersea cable. The array of
optical modules slowly sways back and forth in the ocean currents, leading to 
the necessity of a short-scale active calibration system to record the
geometry for later use in event reconstruction. 
 Sources of noise include bioluminescence, which is reduced by requiring
coincidences of optical modules in a triplet line unit. 
The array was completed in May 2008, with the installation of the final two 
lines. Data from 10 lines, taken over 100 active days of operation from December
2007 to April 2008, showed that upgoing atmospheric neutrinos could be isolated
in the data set. First physics analyses are underway. The sky coverage of ANTARES is
complementary to the south polar detectors, leading to a combined
 full sky coverage. Importantly, ANTARES has a full view of the galactic centre, where
many interesting gamma-ray sources have been observed with HESS. 

There are two other ongoing projects in the Medditeranean - NEMO\cite{NEMO} and NESTOR\cite{NESTOR}.
 Both are still
in the prototyping and construction phase. Together, the three Mediterranean groups have
begun development of a proposal for KM3NET -- a kilometre scale Mediterranean detector\cite{KM3NET}.

\section{Antarctic detectors}

\subsection{Optical Cerenkov detectors: AMANDA and IceCube}

%\subsection{Construction}
The first detection of muon Cherenkov radiation
 in polar ice was made in Greenland in 1990 \cite{muons91}, 
using three photomultipliers deployed to a depth of about 200 metres. 
Following this success, similar tests were made at the South Pole over
the next years, with the AMANDA-A detector deployed in 1993-94 \cite{ice1}. 
Construction of the presently operating 
AMANDA-II detector took place from 1995 to 2000, over which
time 677 optical modules were deployed over 19 strings, to 
depths ranging from 1500 to 2000 metres. The properties of the polar ice, critical
for understanding of the detector, have been measured using light sources
in the array \cite{icepaper}. Although most of AMANDA used analogue signal technology,
 digital technology, eventually chosen for IceCube, was
tested on one string \cite{string18}.

%\subsubsection{Atmospheric neutrinos}
While three neutrino candidates were observed with the first four strings of AMANDA \cite{AMA-B4},
the first compelling evidence of high-energy atmospheric neutrinos came from the 
10 string 1997 data set, where 16 upgoing events were left after data reduction \cite{firstatmos}. Dramatic
improvements in the analysis techniques \cite{recoAMANDA} increased this number to about 300 \cite{atmosnu1997,atmos-nature}.
Over the entire life of AMANDA-II, many thousands of atmospheric neutrinos
have now been observed \cite{ackermann-ps,ps2000-04}. These are the highest energy 
neutrinos ever observed. 
 The observed rate is consistent with the uncertainties in
 theoretical
predictions \cite{bartol2004,honda2004}. A regularised unfolding technique has been used to make a 
best-fit to the originating energy spectrum; again consistency with expectation is 
seen \cite{Munich}. The agreement of the atmospheric neutrino measurements with
expectations shows that the detector is working as expected. 

%\subsubsection{Point sources}
 Several searches for northern hemisphere
point sources of neutrinos have been conducted with the 
AMANDA detector, for the  1997 \cite{ps-1997}, 2000 \cite{ps-2000},
  2000-02 \cite{ps-2000-02} and  2000-04 data sets\cite{ackermann-ps,ps2000-04}.
Several  search methods were used  to look for point sources in the northern sky. 
For each, the expected background for any source is found from off-source data
from the same declination band. The expected sensitivity is found from 
simulations of neutrino interactions, muon propagation, and 
 the full detector response to the Cherenkov light emitted. 
Full-sky searches (looking for a hot spot anywhere in the sky), specific 
source searches, and stacking searches were conducted. 
 The full-sky and specific source searches were optimised in an unbiased fashion
  to produce
the best limit setting potential \cite{mrp}. 
The 90\% confidence level 
sensitivity of the 
full-sky search to an $E^{-2}$ flux (assumed to have a $\nu_{\mu}:\nu_{\tau}$ ratio of 1:1), relatively constant with declination, is about
\esqdnde $<  10^{-10}$ \pointunit. The numbers of observed 
events across the sky were consistent with the background expectations, leading to the
same result for  the average all-sky experimental
limit.  The highest significance seen was 3.7$\sigma$ and, via scrambled
random sky maps, the probability of seeing something this significant
or higher was found to be 69\%. 
Searches for 32 specific candidate sources, and searches made where the events from
 objects belonging to
 common classes were summed, were made. 
Limits were placed on the neutrino fluxes from the objects \cite{Gross-stacking,ps2000-04}. 
 For a source above the horizon, SGR 1806-20, a search for muons from
both neutrinos and gamma-rays was made. With no  significant signal seen,
 limits were placed on the gamma and neutrino fluxes from the source \cite{sgr}.
   While not truly a point source, the galactic plane was searched for an excess of 
neutrinos from cosmic ray interactions with the dust, using similar methods as employed in
 the 
point source searches.
No excess
of events was seen and limits on models were set \cite{Kelley}.
 
Most recently, a new point source analysis was performed on
the full AMANDA-II data set, for the years 2000-06\cite{ps2000-06}.
This work used a full maximum likelihood fit, using the measured angular error of
the reconstructed events and an energy dependent variable in the likelihood
construction to increase the sensitivity
to sources. No evidence for any sources was seen but the improved methods resulted
in better limits than previous angular-binned analyses.

%\subsubsection{GRBs}
Gamma-ray bursts are some of the most energetic phenomena in the universe, with
emission timescales
as short as  seconds. During the life of AMANDA, satellites such as the CGRO, with the
BATSE detector, and the IPN satellites, including HETE and Swift, have recorded
gamma emissions from many GRBs. 
Waxman and Bahcall theorised that GRBs may be the source of the highest energy
cosmic rays \cite{WB97}. In this ``fireball'' model, neutrinos would also be produced. 
The AMANDA data has been searched for neutrinos in spatial and temporal 
coincidence with about 400 GRBs \cite{Kuehn}. The addition of a time cut on the search 
greatly reduces the expected background  to an order of one event over the sum of
all GRBs searched.  No event has been observed in coincidence with
a GRB, consistent with  this small total expected background.  Limits on 
the fluxes from all bursts, classes of bursts,  and individual bursts, have
been placed. The limits from all bursts are within a factor 4 of the Waxman-Bahcall prediction.
 In another analysis, the observations from each individual
 burst are 
interpreted in light of all information known about that burst from other 
wavelengths, via an individually calculated neutrino flux. An
analysis of this type has been 
performed for GRB030329 \cite{Stamatikos}.  The study of further GRBs is in progress. 
 Searches for cascade like events from GRBs  have been made \cite{Hughey}.
 All-time and rolling time
window searches have been performed and limits placed on models of neutrino production.

%\subsubsection{WIMPs}
The mystery of the dark matter, responsible for some 23\% of the 
energy density of the universe, is a target of the search for WIMPs
 (Weakly Interacting Massive Particles) with AMANDA.
A likely dark matter candidate is the neutralino - the lightest 
supersymmetric particle  in most supersymmetric extensions of the 
standard model. 
 After some time, these
would become gravitationally trapped in the centre of the earth and
sun, where they could pair-wise annihilate via several paths to produce
neutrinos. Thus, AMANDA is used in searches for excesses of 
neutrinos from the centre of the earth (1997-99 data \cite{WIMPS97earth,WIMPs97-99earth}),
 and from the sun (2001 data \cite{WIMPs2001sun}).
To date, neither the earth nor sun has been revealed as an annihilation
site for neutralinos, and these non-observations  place bounds on various
parameters in the supersymmetric extensions of the standard model.
 Once
all current data is analysed, these bounds  will be competitive and 
complementary with those from direct 
 detection experiments like CDMS.

%\subsubsection{Diffuse searches}
%\label{diffuse}
To search for a diffuse flux of neutrinos from the sum of sources in the 
universe, one must look for neutrinos in excess of the expectation for
  atmospheric neutrinos. The extra-terrestrial flux is expected to  have
a harder spectrum ($\sim E^{-2}$) than the atmospheric neutrinos ($\sim E^{-3.7}$), 
so searches are designed where event energies are estimated.
 Three types of diffuse search are conducted with AMANDA, one sensitive to muon-neutrinos,
and the other two sensitive to all flavours. The muon search seeks to isolate muon
tracks and use event observables related to the energy.
  One style of all-flavour search focuses on cascade-like events - and is thus
sensitive to neutral and charged current interactions of all flavours.
   Cascades from charged current interactions come from electron and tau neutrinos, and 
from some muon-neutrinos where most of the energy goes into the cascade, leaving only a 
short track from a low energy muon. These searches are mostly sensitive to cascade events
 contained in the detector volume.
  The second type of all-flavour search looks for large cascade and muon events from
extremely high energy neutrino interactions, including events where the cascade or
muon is well outside the volume of the detector. Due to attenuation of neutrinos in the
earth,  these searches are most sensitive to
horizontal events, with the main background being energetic cosmic ray muon bundles.
%  Throughout all these searches, the events energies are estimated using a range of
%techniques. 
  
  Unlike a point source search, a diffuse search strictly has no 
 ``off-source'' region where
data can be used to estimate the background. Thus the analysis relies on 
theoretical predictions of the atmospheric neutrino fluxes for background
estimations. In practice, the observed lower energy events are  used to 
 place some   constraint on the atmospheric models before they are used to estimate
the high energy background. As for other analyses, downgoing muons are used as a 
calibration beam to check that the  detector would be sensitive to the types of
high-energy events expected from extra-terrestrial neutrinos.

  Two \emph{all-flavour cascade} searches have been performed,
 on the 1997 \cite{1997cascade} and
2000 \cite{2000cascade} data sets.
 The limit for the 2000 data  improved by an order of magnitude
over that for 1997. In a similar energy range ($20-5\times10^{4}$ TeV),
 the Baikal collaboration has
recently analysed  1038 days (1998-2003) of 
data from  the NT-200 experiment, leading to a slightly better limit of
 \esqdnde $= 8.1\times10^{-7}$ \diffunit \cite{Baikal98-03}.

 At higher energies, these data sets have been analysed with the  \emph{all-flavour UHE} 
method \cite{1997UHE,Gerhardt-ICRC,Gerhardt-SUSY}.
 Although the sensitivity of the 2000 search ( \esqdnde $= 3.7\times10^{-7}$ \diffunit) was
improved over 1997,  the experimentally obtained limit for 2000 turned out to be the same as that
for 1997, due to the observation of
 a non-significant excess of events.
These limits are the best of any detector at energies up to $\sim 1$PeV.

Searches for a diffuse flux, using reconstructed contained muon events, have been
made on the 1997 \cite{97diffuse}, 2000 and 2000-03 data sets. For the year
 2000 data set, a regularised
unfolding of the energy spectrum was conducted. This spectrum was statistically compared
with the atmospheric neutrino expectation and a limit on a diffuse $E^{-2}$ flux
derived \cite{Munich}. For the 2000-03 data\cite{diffuse2000-03prd}, the muon analysis used the
number of optical module channels per event that reported at least one Cherenkov
photon (\Nch$\!$) as an energy estimator. The harder expected extra-terrestrial flux
would produce a flatter \Nch distribution than that for atmospheric neutrinos.
 Before looking at the data, an optimal cut of \Nch was found in order to produce
the best limit setting sensitivity of the search \cite{mrp,hodgesICRC}.
 The data above this cut (\Nch $> 100$)
were kept blind while the lower \Nch events were compared to atmospheric neutrino 
expectations. The Bartol \cite{bartol2004} and Honda \cite{honda2004}
 atmospheric neutrino fluxes
were varied to account for systematic uncertainties, then constrained by normalisation with the
low \Nch data. The remaining spread in the high \Nch region was used to calculate
an error on the expected number of events above the \Nch $> 100$ cut. 
 Above the cut, 6 events
were seen, where 6.1 were expected. Using the range of atmospheric uncertainty
in the limit calculation \cite{ch} 
leads to a limit on an $E^{-2}$ flux of
muon-neutrinos, at the earth, of
 \esqdnde $= 8.8\times10^{-8}$ \diffunit. This limit is valid in the energy range 16-2500 TeV
 and 
is the best limit
of any neutrino detector to date. Limits were also placed on specific extra-terrestrial
models and on the flux of prompt, charm-meson neutrinos from the earth's atmosphere \cite{hodges-tev}.

%\subsubsection{Supernovae, cosmic ray composition, monopoles and new physics}
AMANDA is a supernova detector, with sensitive coverage of our galaxy \cite{sn-97-98}.
 A burst of low energy electron-neutrinos
from a supernova would produce an increase in the rates of all optical 
modules over a short time ($\sim 10\; {\rm seconds}$).
 The AMANDA supernova system is part of
the Supernova Early Warning System (SNEWS). 
AMANDA, in conjunction with the SPASE surface air shower detector, has been 
used to study the composition of cosmic rays near the knee \cite{composition}. 
Searches for magnetic monopoles have been made, and
Lorentz invariance and 
 decoherence are
 two of the ``new physics'' tests being 
conducted with atmospheric neutrino data from AMANDA.

%\subsubsection{Construction and Performance}
The first of the next generation kilometre scale neutrino
telescopes, IceCube\cite{icecube}, will consist of an in-ice cubic kilometre
neutrino detector, and a kilometre square surface cosmic
ray air shower detector (IceTop).
%\subsection{Construction and Performance}
  Construction began
 at the South Pole 
during the austral summer 2004-05, with 1 in-ice string, and
4 IceTop stations deployed \cite{icperf}.
The goal is to complete construction in early 2011, with
80 strings (4800 modules) and stations (320 modules)
 completed. The in-ice strings will
instrument a kilometre volume between 1500 and 2500 metres
depth, and the IceTop array will cover a square kilometre
at the surface. The same design of DOM (Digital Optical 
Module) is used throughout the detector. These consist of
pressure spheres containing 10 inch photomultiplier tubes,
the signals of which are digitised inside the module and 
then sent to the surface data acquisition system. The DOMs
differ from the AMANDA modules in that the full time series
of photons (the ``waveform'') is captured. 

The holes are drilled with a hot water system, 
taking about 30 hours to drill to the final depth, then
10 hours to
 ream back up, depositing more energy to leave a hole at
the correct size during the string deployment.
Deployment of a string now takes about 4-9 hours - 3-7 hours for
module attachment,  then 1-2 hours to lower to the final depth. 
IceTop tanks are installed in shallow trenches dug near each
string location, and are filled with water, which is 
allowed to slowly freeze back about the modules, to prevent 
formation of bubbles. 

%\subsection{Performance to date}
The deployed hardware has performed up to expectations
to date. Detailed studies of the first string and IceTop
tank behaviour have been published \cite{icperf}. Even with
one string, upward moving
events were detected, consistent
with an atmospheric neutrino origin.
After this initial season, subsequent seasons have seen 8, 13 and 18 strings
deployed for a current total of 40 strings. 
 Air showers have been reconstructed
with IceTop, and coincident events, where IceTop sees an air shower and
the in-ice array sees the penetrating muons, have been studied. 
The 40 string array has been running for more than six months and data
analysis has started. Analyses of the 9 and 22 string data have been 
completed, with no evidence of any sources seen. 

%\subsubsection{Physics potential}
An initial potential performance study for the 
in-ice array of IceCube was completed before
construction began \cite{icsens}. The simulation and reconstruction 
programs were those used in AMANDA,
adapted to  the larger IceCube detector. 
As such, no usage of the DOM waveform information was made  in
the reconstruction. The assumed flux of charm atmospheric
 neutrinos \cite{rqpm}
was chosen conservatively; if in reality this background
 turns out
smaller, then the predicted sensitivities will be better than
 those quoted. 
 A median angular resolution of better than 1$^\circ$
is seen for muon energies greater than 1 TeV. The effective area for
muon detection exceeds the geometric kilometre area at 10 TeV, rising
to 1.4 square kilometres for events in the 1 to 100 PeV energy range.
The sensitivity to diffuse and point sources of neutrinos has been
estimated. For three to five years of observation, the limit on an $E^{-2}$
flux of diffuse neutrinos would be 
about thirty times smaller
than the AMANDA-II four-year muon limit,
 and a flux one-tenth of the AMANDA-II
limit would be detectable at $5\sigma$ significance in that time. 
For point sources, similar results are obtained. For GRBs, the
Waxman-Bahcall flux would be constrained after the observation of
about 100 GRBs, and 500 GRBs would be needed to observe that flux at
a $5\sigma$ significance.

\subsection{Radio-Cerenkov detectors: RICE and ANITA}
At extremely high energies, the interaction of an ultra-high energy neutrino with the
antarctic ice can produce coherent Cerenkov light in the radio frequency range. 
 The first detector to exploit this principle was RICE (Radio Ice Cerenkov Experiment)
which was deployed at the south pole. RICE consists of radio receivers which are
deployed to shallow depths in some of the AMANDA holes. No events consistent with
a neutrino interaction have been seen and thus limits have been placed on the
expected numbers of such neutrinos.

The ANITA experiment\cite{ANITA} takes the radio concept into the skies above Antarctic. Radio receivers are
flown on a baloon up to about 100000 feet, from which a large volume of ice is observable.
    Two flights have occurred so far: the test flight of the smaller-scale
 ANITA-lite\cite{ANITA-lite} and then a full ANITA mission\cite{ANITA-nu2008}, 
launched  December 15th, 2006 and flown for 35 days. The analysis of these data has been 
completed. Whilst there no physics backgrounds at the sensitive energy range of the 
experiment, there are many possible manmade backgrounds across the constinent which must be 
eliminated.  After analysis of the data, no candidate neutrino events were found. 
Limits were placed on the fluxes of neutrinos at the highest energies. 

The radio technique is being further pursued in Antarctica: there will be another ANITA flight and 
studies are underway for further in-ice detectors. Development work is underway for large
arrays of buried-surface or shallow-hole receivers on the Ross ice shelf,
 or spread over a large area centred on the IceCube
detector.

\section{Conclusions}
  The long-held dream of a large volume, high energy neutrino detector is finally
 a reality at several sites around the world. The last decade has seen  great progress
in
technology, deployment, and analysis technique 
development 
for these detectors.
 The Lake Baikal, ANTARES, AMANDA, RICE, IceCube and ANITA detectors are
operational and producing physics data. 
These detectors  have unprecedented sensitivity to sources
of extra-terrestrial neutrinos, which will  hopefully lead to new discoveries about the 
nature of the cosmos.


\begin{thebibliography}{99}
                                                                                                                    

\bibitem{auger}
http://www.auger.org


\bibitem{auger-science}
Auger collaboration 2007 {\it Science} {\bf318} 938

\bibitem{milagro}
www.lanl.gov

\bibitem{Milagro-Cygnus}
Abdo et al, {\it Ap J Lett}, {\bf664}, L91 (2007)


\bibitem{HESS}
http://www.mpi-hd.mpg.de/hfm/HESS/HESS.html

\bibitem{MAGIC}
http://wwwmagic.mppmu.mpg.de

\bibitem{VERITAS}
http://veritas.sao.arizona.edu

\bibitem{TeVCat}
http://tevcat.uchicago.edu



%\bibitem{reines}
%K.~Greisen, {\it Ann.\ Rev.\ Nucl.\ Part.\ Sci.\ }  {\bf 10}, 63 (1960), M.~A.~Markov in {\it Proceedings of the 1960 International Conference on High Energy Physics}, E.C.G. Sudarshan, J.H.Tinlot and A.C.Melissinos Editors, 578 (1960).
%


\bibitem{PR}
Gaisser~T~K, Halzen~F and Stanev~T 1995 {\it Phys.\ Rept.} {\bf 258} 173 
[{\it Erratum} 1995 {\bf 271} 355 ], hep-ph/9410384; Learned~J~G and Mannheim~K 2000
 {\it Ann. Rev. Nucl.
Part. Science} {\bf 50} 679; Halzen~F and Hooper~D 2002 {\it Rept.\ Prog.\  Phys.} {\bf65} 1025, arXiv:astro-ph/0204527;
Halzen~F 2006  {\it Proc. of 'The multi-messenger approach to high-energy gamma ray sources,'} Barcelona 

\bibitem{Baikal98-03}
 Dzhilkibaev Z \ea 2006
{\it Astropart. Phys.} {\bf25}  140

\bibitem{ANTARES}
http://antares.in2p3.fr/

\bibitem{ANTARES-nu2008}
Carr~J, for the ANTARES collaboration, Neutrino 2008, Christchurch, May 2008

\bibitem{NEMO} 
http://nemoweb.lns.infn.it/project.html

\bibitem{NESTOR} 
http://www.nestor.noa.gr

\bibitem{KM3NET}
http://www.km3net.org/home.php

\bibitem{muons91}
   Lowder~D~M,  Miller~T,  Price~P~B,  Westphal~A,  Barwick~S~W,  Halzen~F and  Morse~R 1991
  {\it Nature} {\bf353}  331

\bibitem{ice1} %42
     Askebjer~P {\it et al} 1995
     {\it Science} {\bf267}  1147-1150

\bibitem{icepaper}
\am 2006
 {\it Journal of Geophysical Research}
   {\bf 111}  D13203

\bibitem{string18}
\am 2006
{\it Nucl. Inst. and Meth. in Phys. Res.}, A. {\bf 556} 169
                                                                                                                   


\bibitem{AMA-B4}
 Andres E {\it et al} 2000
{\it Astropart. Phys.} {\bf 13}  1


\bibitem{firstatmos}
 Karle A, for the AMANDA collaboration 1999
 {\it Proc.  26th ICRC, Salt Lake City, Utah} 
                                                                                                                                         
\bibitem{recoAMANDA}
\ar 2004
{\it Nuclear Instruments and Methods in Physics Research} A {\bf 524}  169
                                                                                                                                         
\bibitem{atmosnu1997}
\ar 2002
{\it Phys. Rev. D} {\bf66}, 012005 
                                                                                                                                         
                                                                                                                                         
                                                                                                                                         
\bibitem{atmos-nature}
 Andres E \ea 2001
{\it Nature} {\bf410} 441 

\bibitem{ackermann-ps}
 Ackermann M, for the IceCube Collaboration 2006 
{\it Proc. of 'The multi-messenger approach to high-energy gamma ray sources,'} Barcelona 
                                                                                                                                         

\bibitem{ps2000-04}
\ab 2007
{\it Phys. Rev. D} {\bf75} 102001


\bibitem{bartol2004}
    Barr~G~D, Gaisser~T~K, Lipari~P, Robbins~S and  Stanev~T 2004
{\it Phys. Rev. D} {\bf70}, 023006 
                                                                                                                                         
\bibitem{honda2004}
     Honda M,  Kajita T, Kasahara K and Midorikawa S 2004 {\it Phys. Rev.} D {\bf70}, 043008 


\bibitem{Munich} K.~M\"unich for the IceCube Collaboration 2005
  \puneicrc



\bibitem{ps-1997} \ar 2003
{\it Astrophys. J} {\bf583} 1040 
                                                                                                                                         
                                                                                                                                         
\bibitem{ps-2000}
\ar 2004
{\it Phys. Rev. Lett.} {\bf92} 071102 
                                                                                                                                         
\bibitem{ps-2000-02} \am
2005
{\it Phys. Rev. D} {\bf71} 077102 


\bibitem{mrp}
 Hill~G~C and  Rawlins~K 2003
{\it Astropart. Phys.} {\bf19}  393

\bibitem{Gross-stacking} \ab 2006 {\it Astroparticle Physics}, accepted 


\bibitem{sgr}
\ab 2006
{\it Phys. Rev. Lett.}  {\bf97} 221101

\bibitem{Kelley}  Kelley~J~L for the IceCube Collaboration
 \puneicrc


\bibitem{ps2000-06}
AMANDA collaboration, http://arxiv.org/abs/0809.1646, 2008

\bibitem{WB97}
 Waxman E and Bahcall J 1997
{\it Phys. Rev. Lett.} {\bf78} 2292-2295 
                                                                                                                                         
\bibitem{Kuehn} Kuehn K for the IceCube Collaboration and the IPN Collaboration
 \puneicrc

\bibitem{Stamatikos} Stamatikos M, Kurtzweil J and Clarke M~J for the IceCube Collaboration
({\it Preprint:} astro-ph/0510336)

\bibitem{Hughey} Hughey B and Taboada I for the IceCube Collaboration,
  \puneicrc

\bibitem{WIMPS97earth}
\ar 2002
{\it Phys. Rev.} D {\bf66} 032006 
                                                                                                                                         
\bibitem{WIMPs97-99earth}
\ab 2006
   {\it Astropart. Phys.} in press 
                                                                                                                                         
                                                                                                                                         
%\bibitem{Hubert} D.~Hubert, A.~Davour, C.~de los Heros for the IceCube Collaboration, {\it Search for neutralino dark matter with the AMANDA neutrino detector}
%, these proceedings
                                                                                                                                         
\bibitem{WIMPs2001sun}
\ab 2006
{\it Astropart. Phys.}   {\bf24} 456-466 

\bibitem{1997cascade}
\ar 2003
{\it Phys. Rev.} D {\bf67} 012003 
                                                                                                                                         
\bibitem{2000cascade} \am 2004
 {\it Astropart. Phys.} {\bf22} 127 






\bibitem{1997UHE}
\am 2005
{\it Astropart. Phys.} {\bf22} 339 

\bibitem{Gerhardt-ICRC} Gerhardt L for the IceCube Collaboration, \puneicrc
                                                                                                                                         
\bibitem{Gerhardt-SUSY}  Gerhardt L for the IceCube Collaboration 2006 Proc. SUSY06 

\bibitem{97diffuse}
\ar 2003
{\it Phys. Rev. Lett.} {\bf90} 251101 
                                                                                                                                         
\bibitem{diffuse2000-03prd}

 Achterberg A {\em et al.}  (IceCube collaboration), 
  Phys. Rev. D {\bf76}, 042008 (2007); erratum, \emph{ibid} {\bf77}, 089904(E) (2008)


                                                                                                                                         
\bibitem{hodgesICRC} Hodges~J  for the IceCube Collaboration,  \puneicrc


%\bibitem{ch} %17
% Cousins~R~D and  Highland~V~L 1992 {\it Nucl. Ins. Meth. Phys. Res.} {\bf A320} 331 (1992)
%\bibitem{fc}
%Feldman G and Cousins R 1998 {\it Phys. Rev.} D {\bf 57} 3873
%\bibitem{conrad} %18
% Conrad J,  Botner O,  Hallgren A and de los Heros C 2003
% {\it Phys. Rev.} D {\bf 67}, 012002 
%\bibitem{gch-sys} %19
% Hill G~C 2003 {\it Phys. Rev.} D 118101 


\bibitem{ch} %17
 Cousins~R~D and  Highland~V~L 1992 {\it Nucl. Ins. Meth. Phys. Res.} {\bf A320} 331, 
Feldman G and Cousins R 1998 {\it Phys. Rev.} D {\bf 57} 3873,
 Conrad J,  Botner O,  Hallgren A and de los Heros C 2003
 {\it Phys. Rev.} D {\bf 67} 012002,
 Hill G~C 2003 {\it Phys. Rev.} D 118101






 
\bibitem{hodges-tev} Hodges J for the IceCube Collaboration 2006 {\it Proc.
TeV Particle Astrophysics II,} Madison, to appear in {\it J. Phys.: Conf. Series}
                                                                                                                                         
\bibitem{sn-97-98}
\ar 2002
{\it Astropart. Phys.} {\bf16}  345
                                                                                                                   


\bibitem{composition}
\ar (AMANDA and SPASE Collaborations) 2004
{\it Astropart. Phys.}  {\bf21}  565
                                                                                                                                         

\bibitem{icecube}
http://icecube.wisc.edu  

\bibitem{icperf}
\ab 2006
{\it Astropart. Phys.}  {\bf26} 155
                                                                                                                                         



\bibitem{icsens}
\ar 2004
{\it Astropart. Phys.} {\bf20} 507
                                                                                                                                         

\bibitem{rqpm}
Bugaev~E~V \ea 1998 {\it Phys. Rev. D} {\bf58}  054001


\bibitem{ANITA}
http://www.phys.hawaii.edu/~anita/web/index.html

\bibitem{ANITA-lite}
 Barwick~S~W et al, Phys Rev D, {\bf96}, 2006, 171101

\bibitem{ANITA-nu2008}
 Gorham~P~W for the ANITA collaboration, Neutrino 2008, Christchurch, May 2008

                                                                                                                    
                                                                                                                    
\end{thebibliography}
\end{document}